\documentclass[aps,prl,groupedaddress,nofootinbib,amsmath,amssymb,reprint]{revtex4-1}

\usepackage{graphicx}
\usepackage{bm}

\usepackage{xcolor} \usepackage{ulem} \usepackage{changebar}
\colorlet{ins}{blue} \colorlet{del}{red}

\newcommand\TLSdel[1]{\cbdelete{}\textcolor{del}{\sout{#1}}}

\newcommand{\atomstfs}[4]{\ensuremath{|#1#2_{#3},m_J=#4\rangle}}

\newcommand{\atomsthfs}[5]{\ensuremath{|#1#2_{#3},F=#4,m_F=#5\rangle}}

\newcommand{\gsta}{\atomsthfs{5}{S}{1/2}{2}{0}}

\newcommand{\sta}{\ensuremath{|a\rangle}}
\newcommand{\stb}{\ensuremath{|b\rangle}}
\newcommand{\str}{\ensuremath{|r\rangle}}

\usepackage{graphicx}
\begin{document}


\title{Atomic Fock State Preparation Using Rydberg Blockade}

\author{{Matthew Ebert}}%
 \email{mebert@wisc.edu}
\author{{Alexander Gill}}
\author{{Michael Gibbons}}%
\author{Xianli Zhang}%
\author{{Mark Saffman}}%
\author{{Thad G. Walker}}%
\affiliation{%
Department of Physics, University of Wisconsin, 1150 University Avenue, Madison, Wisconsin 53706, USA    
}%

\date{\today}

\begin{abstract}
  We use coherent excitation of 3-16 atom ensembles to demonstrate collective Rabi flopping mediated by Rydberg blockade.  Using calibrated atom number measurements, we quantitatively confirm the expected $\sqrt{N}$ Rabi frequency enhancement to within {4\%}.  The resulting atom number distributions are consistent with essentially perfect blockade.
 We then use collective Rabi $\pi$ pulses to produce ${\cal N}=1,2$ atom number Fock states with fidelities of 62\% and 48\% respectively.  The ${\cal N}=2$ Fock state shows the collective Rabi frequency enhancement without corruption from atom number fluctuations. 
  \end{abstract}
\maketitle


Ensembles of cold neutral atoms localized in micron-sized clouds interact collectively with laser light tuned to excite $n\sim 100$ Rydberg states.
Within such clouds the interactions between two or more Rydberg atoms are many orders of magnitude greater than the interactions between ground-state atoms.
Thus while a single photonic absorption is resonant, subsequent photonic absorptions are made off-resonant by  Rydberg-Rydberg interactions.
For sufficiently cold atoms, this "blockade" energetically constrains the $N$ atom ensemble to an effective 2-level Hilbert space consisting of either $N$ ground-state atoms or $N-1$ ground state atoms and 1 Rydberg excitation.
The sharing of the excitation between the $N$ atoms causes atom-light couplings to be collectively enhanced by $\sqrt{N}$ \cite{Lukin01,Saffman10}.

For $N=2$, Rydberg blockade \cite{Urban2009,Gaetan2009} has been exploited to produce entanglement \cite{Wilk2010,Isenhower2010,Zhang2010} and to observe $\sqrt{2}$ Rabi enhancement \cite{Gaetan2009}.  Collective Rabi oscillations at large $N$ were also recently observed \cite{Dudin2012}.
{When the cloud size allows multiple Rydberg excitations, blockade  results in excitation suppression and  dramatically increased optical non-linearities.  This works even at the single photon level, \cite{Peyronel2012,Dudin2012a,Maxwell2013,Honer2011} and allows entanglement of light and atomic excitations\cite{Li2013}.} 

\begin{figure}[hbt]
  \includegraphics{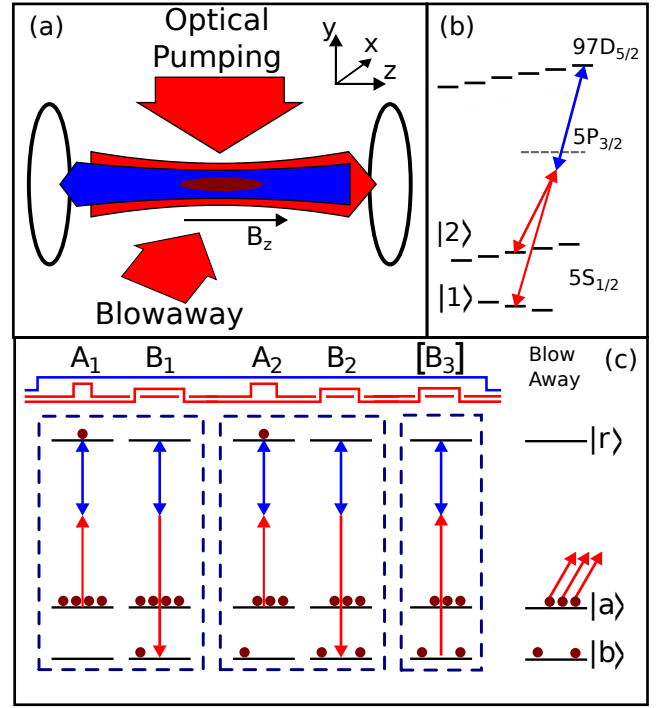}
  \caption{(Color online)
    (a) {Experimental geometry.
      Counter-propagating 780 and 480 nm Rydberg excitation lasers, parallel to an applied magnetic field,  couple \sta{} or \stb{} to \str{}.
    Optical pumping and state selective blow away beams are incident on the ensemble in the perpendicular plane.}
    (b) Two-photon excitation diagram.
    (c) Fock state generation pulse sequences.
    Sequential pairs of  A and B excitation pulses perform population transfer from $\sta \rightarrow \str \rightarrow \stb$.
    Ideally, the Rydberg blockade mechanism moves a single atom to \stb{} per $A-B$  pulse pair.
  After two $A-B$ pulse pairs, the $B_3$ pulse optionally probes 2-atom Fock state dynamics.}
  \label{fig_maindia}
\end{figure}

The classic signature of coherent collective behavior is collective Rabi flopping, as  emphasized in the original Lukin{ \it et al.} proposal \cite{Lukin01}.  Fluctuations in atom loading statistics produce, through the $\sqrt{N}$ dependence, inhomogeneous broadening that dephases the collective Rabi manipulations.  This is important, for example, 
for potential use in collective quantum gates \cite{Beterov2013},  protocols for deterministic single photon sources \cite{Saffman2002}, or entanglement of single-atom and collective qubits \cite{Saffman2005b}.
To minimize  this effect, one would like to be able to reduce the atom number fluctuations below the classical Poissonian limit, also proportional to $\sqrt{N}$.

In this Letter we experimentally realize a collective protocol \cite{Saffman2002} for deterministic production of single and two-atom Fock states.
We load an ensemble of $3<\bar{N}<16$ atoms into a single dipole trap and extract single atoms via collective Rabi $\pi$-pulses between one ground-state hyperfine level and a Rydberg state, followed by stimulated emission into a second ground hyperfine level{, Fig. \ref{fig_maindia}(c)}.
{We quantitatively verify the $\sqrt{N}$ enhancement factor with a precision of $4\%$.}
Subsequent sequences of such pulse pairs allows production of  multi-atom Fock states.
We demonstrate sub-Poissonian production of single and two-atom Fock states using this method. 

Our basic apparatus is quite similar to our previous work \cite{Isenhower2010,Zhang2010}.
Indeed, the collective Rabi flopping protocols are similar to protocols for single-atom qubit control, and hence are convenient for loading  arrays for neutral atom quantum computing.
We transfer a small number of Rb atoms (up to ~30)  from a magneto-optical trap into a 1.5 mK deep 1.06 $\mu$m far-off resonance trap (FORT) focussed to a waist of 3.0 $\mu$m.
The atoms are then laser-cooled to {100-150} $\mu$K, during which time light-assisted collisions induced by the cooling light cause atoms to be ejected from the FORT.
Varying the cooling time allows us to  realize a mean atom number, $\bar{N}$, from 0.5 to 16 atoms.  Measurement of $\bar N$ will be discussed later.
The spatial distribution of the trapped atoms is a 7 $\mu$m long and {$< 0.5\mu$m} wide Gaussian distribution oriented along the FORT propagation direction.  Calculations indicate that the Rydberg-Rydberg interaction  \cite{Walker2008} is  11 MHz at a typical 12 $\mu$m atom-atom distance, sufficient for the 1 MHz scale Rabi flopping studied here.
Once trapped, the atoms are optically pumped into the \gsta{} clock state, and the FORT is turned off for 3-6 $\mu s$ while the Fock state pulse sequence is applied.

The Fock-state pulse geometry is shown in Fig. \ref{fig_maindia}(a).
We perform independent coherent two-photon excitations between either of the two ground states $(|a\rangle,|b\rangle)=|5S_{1/2},F=(2,1),m_F=0\rangle$ and the \str{}=\atomstfs{97}{D}{5/2}{5/2} Rydberg state.  
Two independently switchable 780 nm lasers, with waists of $\omega_{(x,y)}=(9,7)$  $\mu$m, energetically select the hyperfine level  coupled to \str{}.
{Both of} these lasers co-propagate with the FORT laser.
A counter-propagating 480 nm beam {with waists of $\omega_{(x,y)}=(5.6,4.7)$ $\mu$m}, which is left on continuously, provides the second step to the Rydberg state.  Each excitation laser is  locked to a different mode of a high finesse cavity.  The single-photon Rabi frequencies for the 2-photon transition are typically $(\Omega_{480}, \Omega_{780})=2\pi\times (17, 160)$ MHz,  with a -2.1 GHz detuning from the 5P$_{3/2}$ level, giving a single-atom two-photon Rabi frequency of $\Omega_1=2\pi\times 750$ kHz.
The timing of each pulse is controlled by the duration of  the respective 780 nm beams.
In the following, we refer to a Rabi oscillation between (\sta{},\stb{}) and \str{} as an  $(A_p,B_p)$ pulse, where $p$ refers to the pulse sequence number in Figure \ref{fig_maindia}(c).
All pulse times $t$ are chosen to have pulse area $\theta={\pi}= \sqrt{\bar N}\Omega t$, unless explicitly noted.  

Following the Fock state pulses, the number ($N_b=0, 1$, or 2) of atoms in state \stb{} is determined by first ejecting atoms from \sta{} using resonant light \cite{Isenhower2010}, then collecting light from the remaining atoms while laser cooling for 20 ms.
Measurements show {that atoms in \sta{} can be ejected with a fidelity of 97\%}.
Atoms remaining in Rydberg states at the end of a pulse sequence leave the trap after the FORT is turned back on \cite{Urban2009}, so population in Rydberg states is not {directly} detected in this experiment.

\begin{figure}[bh]
  \includegraphics[width=\linewidth]{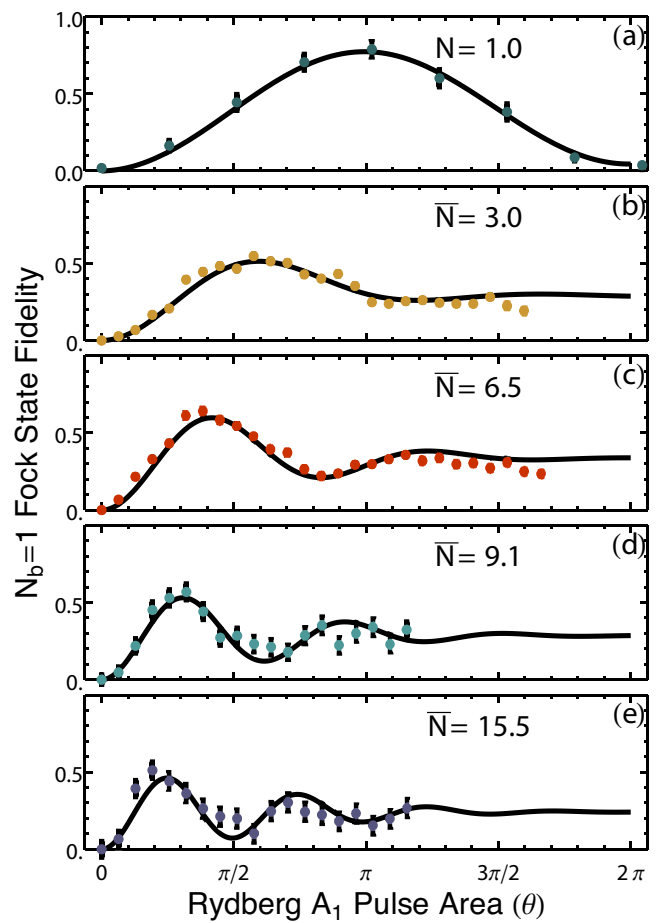}
  \caption{\label{fig_collosc}
    (Color online)
    Rabi oscillations between \sta{} and \str{} for various atom number distributions.
    The single atom detection probability is shown as a function of the pulse area {$\theta=\Omega_{1}t$} of the Rydberg A excitation.
    (a) The first $2\pi$ rotation for exactly one atom.
    (b-e) The \stb{} populations show an atom number dependent frequency for ensemble means of, respectively, $\bar{n}=3.0,6.5,9.1,15.5$. 
    The solid black lines are the fits to Eq. (\ref{eqn_collosc}).  
  }
\end{figure}

Beginning with an {$\bar{N}$} atom ensemble initialized in \sta{}, the  $A_1(\theta)$ pulse produces a collective Rabi oscillation between the state $|g\rangle=|a_{1}a_{2}...a_{N}\rangle$ and the  symmetric singly-excited {W state}  $|r\rangle=N^{-1/2}\sum_{i=1}^{N}{|a_{1}a_{2}...r_{i}...a_{N}\rangle}$.
The  $B_1$ pulse, calibrated using single-atom Rabi oscillations out of state \stb, then drives a single-atom $\pi$ pulse between the single Rydberg atom and the unpopulated \stb{} state.  Ideally, this sequence should produce a single-atom Fock state in \stb.

\begin{figure}[htp]
  \centering
  \includegraphics[width=\linewidth]{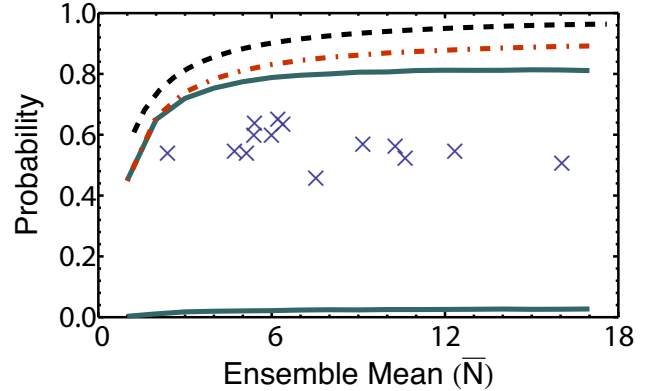}
  \caption{
    (Color online) ${\cal N}=1$ Fock state production fidelity as a function of mean ensemble number, and quantum Monte Carlo simulations.  The black dashed line assumes ideal blockade and perfect excitation conditions.  The red dot-dashed line adds in realistic experimental imperfections, and the green solid line includes finite blockade strength.  The solid green line at the bottom shows the predicted 2-atom production.
  }
  \label{fig_broadeningfull}
\end{figure}

Figure~\ref{fig_collosc} shows the results of  measurements of  $N_b$ after  $A_{1}(\theta) B_1$ sequences.  As the number of atoms is successively increased from 1 (top) to 15.5 (bottom), the Rabi frequency {increases} as expected from collective enhancement.
We fit each data set to the following model for the probability $p_1(t)$ for one atom to be in \stb{} as a function of $A_1(\theta)$ pulse time:
\begin{equation}
  \label{eqn_collosc}
  p_1(t)=\frac{\epsilon}{2}\sum_{N=0}^{N_{max}}P_{\bar{N}}(N)\left[1-\cos\left(\sqrt{N}\Omega_1 t\right)e^{-t/\tau}\right],
\end{equation}
where $P_{\bar{N}}(N)$ is the Poisson distribution of initial atom numbers, and $\tau$ is the decay time of the Rabi oscillations.
Both $\Omega_1$ and $\tau$ (typically $2\pi\times$750 kHz and 5 $\mu$s) are measured from  single-atom Rabi flopping. 
A two parameter fit for each data set  returns the  mean atom number $\bar N$ and an overall scaling factor $\epsilon$, to be discussed with Fig~\ref{fig_broadeningfull}.

 We separately measure $\bar N$ by collecting fluorescence scattered by the atoms from short (3 ms) pulses of cooling light. In the $>10^{11}$ cm$^{-3}$ density cloud, the calibration of number of scattered photons per atom is affected by significant light-assisted collision loss.  In separate experiments we {measure} the relevant two-body loss rates and \TLSdel{}{implement} the relevant calibrations \cite{SM}.
 
\begin{figure}[tb]
  \includegraphics[width=\linewidth]{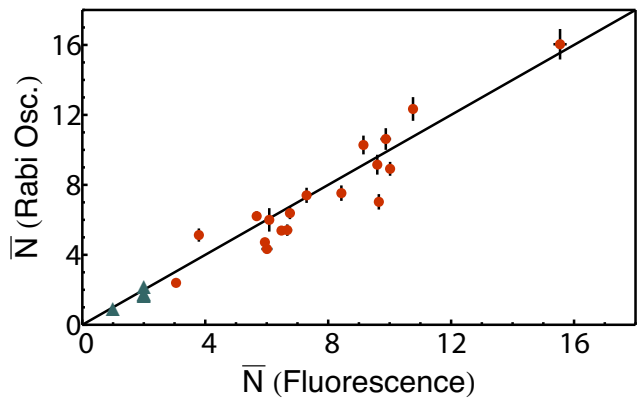}
  \caption{\label{fig_collenc} (Color online) Mean number of atoms in the ensemble, as deduced from collective Rabi oscillations (ordinate), and by fluorescence (abscissa).
    {The red circles} are data from Poisson-distributed atom ensembles, the green triangles have exactly 1 or 2 atoms.
    The solid black line, of slope 1, shows that the collective oscillation frequency closely follows the predicted $\sqrt{N}$ dependence.    }
\end{figure}

A comparison between $\bar N$ as deduced from the collective Rabi oscillations, and from the direct atom number measurements, is shown  in Fig.~\ref{fig_collenc}, along with a line of slope 1.
The close agreement quantitatively confirms the phenomenon of collective Rabi frequency enhancement in the strong blockade limit.    Allowing the slope to vary gives a best fit of 0.96.

Fig. \ref{fig_collosc} demonstrates that when the Rydberg $A_1$ pulse area is chosen to be a collective $\pi$-pulse, i.e. $\Omega_{\bar{N}}t=\pi$, and the Rydberg $B_1$ pulse is set to a single atom $\pi$-pulse, our procedure is capable of creating an {${\cal N}=1$} Fock state in which a single atom of the ensemble has been transferred to the state \stb{} with an efficiency as high as 63.3\%.
The observed distribution of $N_b$ is 35\% $N_b=0$, 63.3\% $N_b=1$, and 1.3\% $N_b=2$.
The number of $N_b=2$ cases  observed is  consistent with our known efficiency for ejecting the atoms {in \sta{}}, implying that any double Rydberg excitations due to imperfect blockade do not transfer to \stb{}.
The resulting Fock state distribution is very sub-Poissonian with a Mandel parameter $Q=\sigma_{N_b}^2/\bar{N_b}-1=-0.62\pm 0.03$.  

The ${\cal N}=1$ Fock state preparation procedure can  be generalized to ${\cal N}>1$ by simply repeating the Rydberg $A$ and $B$ pulse sequence ${\cal N}$ times, with each pulse area set to be a collective $\pi$-pulse for the number of coupled atoms.  Thus the ideal collective Rabi frequencies for pulses $A_2$ and $B_2$ are $\sqrt{\bar N-1}\Omega_1$ and $\sqrt{2}\Omega_1$.
To study ${\cal N}=2$ Fock state preparation, we first consider the possible outcomes of an $A_1B_1$ pulse sequence followed by an $A_2$ pulse. There are four cases $|N_r,N_b\rangle$:
  \medskip
 
\parbox{3.0 in}{  1. neither pulse sequence succeeds: $|0,0\rangle$}\smallskip
  
\parbox{3.0 in}{  2. $A_1B_1$ succeeds, but $A_2$ fails,  resulting in one atom in \stb{}: $|0,1\rangle$}\smallskip
  
 \parbox{3.0 in}{ 3. $A_1$ fails, and $A_2$ succeeds, resulting in one Rydberg atom: $|1,0\rangle$}\smallskip

\parbox{3.0 in}{  4.  All pulses succeed, resulting in one atom in both \stb{} and \str{}: $|1,1\rangle$}\medskip\\
 Finally, a fourth pulse, $B_2(\theta)$, couples $\stb{} \leftrightarrow \str{}$ and evokes different behavior for each possible state mentioned.
For $|0,0\rangle$, $B_2$ has no effect. Both single-atom outcomes  oscillate between $|0,1\rangle \leftrightarrow |1,0\rangle$ at the single atom Rabi frequency but are out of phase with each other.
The $|1,1\rangle$ state, however, uniquely oscillates between $|1,1\rangle \leftrightarrow |0,2\rangle$ at a $\sqrt{2}\Omega_1$ enhanced Rabi frequency.  Blockade forbids population of $|2,0\rangle$.
The  population of the \stb{} state will then be either 0, 1, or 2 atoms.

\begin{figure}[htp]
\centering
\includegraphics[width=\linewidth]{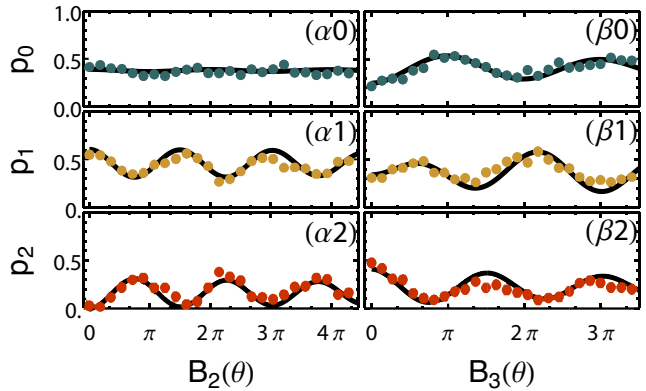}
\caption{
  (Color online)
  ($\alpha$) Evolution of $N_b$=0, 1 and 2 atom  populations using the $A_1B_1A_2B_2(\theta)$ protocol.
  ($\beta$) Output of the ${\cal N}=2$ Fock state production 
   as a function of $B_3$ area.
}
\label{fig_fsosc}
\end{figure}

Figure~\ref{fig_fsosc}($\alpha$) shows the probabilities for observing 0, 1, or 2 atoms in state \stb{} after the $A_1B_1A_2B_2(\theta)$ sequence.
As expected, the probability of observing two atoms, Fig.~\ref{fig_fsosc}($\alpha$2), begins at 0 and oscillates at $\sqrt{2}\Omega_1$.
{Note that the decay time of the two atom collective oscillation is set by the same decoherence processes as would be observed in single atom-atom Rabi oscillations, and there is no additional dephasing from atom number fluctuations because exactly two atoms participate in the oscillation.}
The $N_b=0$ signal, Fig.~\ref{fig_fsosc}($\alpha$0), potentially has contributions from the single atoms states $|0,1\rangle$ and $|1,0\rangle$, but these oscillations are out of phase and cancel if the populations of those states are equal, as is {nearly} the case for this data.

The $N_b=1$ signal, Fig.~\ref{fig_fsosc}($\alpha$1), potentially has contributions from Rabi oscillations of the states $(|1,0\rangle,|0,1\rangle,|1,1\rangle)$ at frequencies $(1,1,\sqrt{2})\Omega_1$.
Again, the first two are cancelled if their populations are equal, leaving only the collective 2-atom signal.

The probability of producing the Fock state $|0,2\rangle$ is 32\% for this data, for which the FORT drop time was extended to 6.34 $\mu$s to see 3 full collective Rabi oscillations.
As a result,   additional high velocity atoms  are not recaptured when the FORT is restored, reducing the signal size.
For 2 $\mu$s FORT drops, we have observed up to $48\pm2\%$ $N_b=2$.  For example, the state produced at the beginning of Fig.~\ref{fig_fsosc}($\beta$) has $Q=-0.50\pm0.05$.

The full $F=2$ Fock sequence can be further probed by restoring the FORT for 0.5 ms, enough to eject {any Rydberg population from} $|1,0\rangle$ and $|1,1\rangle$, effectively leaving only  ground-state populations in the states $|0,0\rangle, |0,1\rangle$, and $|0,2\rangle$.
This removes the cancellation between the $|0,1\rangle$ and $|1,0\rangle$ signals {observed} in Fig.~\ref{fig_fsosc}($\alpha$).
Now, as shown in Fig.~\ref{fig_fsosc}($\beta$), oscillations at $\Omega_1$ are observed in the $N_b=0$ data, and the $N_b=1$ data have both $\Omega_1$ and $\sqrt{2}\Omega_1$ signals superposed.
{The fits to the oscillations have only the 3 initial state populations as adjustable parameters, and assume no  atom number fluctuations.}

Both the $N_b=1$ and $N_b=2$ Fock state populations are consistent with a single collective $AB$ sequence success probability of {$0.65-0.70$}.
In Fig.~\ref{fig_broadeningfull} we show the ${\cal N}=1$ Fock state population as a function of $\bar N$.
We also show  the results of a quantum Monte Carlo model of a collective $A_1B_1$ pulse sequence.
The model considers the known significant
 sources of experimental imperfections, which include Doppler shifts, the distribution of AC-Stark shifts and Rabi frequencies from the Gaussian intensity distributions, spontaneous emission from the intermediate 5p state (spontaneous emission from the Rydberg states is negligible on 5 $\mu$s timescales), and 1 $\mu$m misalignments of the excitation lasers.  For a single atom, the model reproduces our observed AB success probability of 85\%.  For multiple atoms, the model allows both single and double Rydberg excitations.  The double excitations  during the A portion of the sequence primarily consist of atom pairs at extreme ends of the cloud.  Those  double excitations still experience some Rydberg-Rydberg interactions, and therefore are  off-resonant for the B deexcitation portion of the sequence, thereby suppressing the number of occurrences of $N_b=2$.  The lines in Fig.~\ref{fig_broadeningfull} show, from highest to lowest, the predicted fidelity assuming perfect Rabi flopping and infinite blockade;  experimental imperfections with infinite blockade; and finally including both experimental imperfections and finite blockade.  The 
black dashed curve corresponds to the fit parameter $\epsilon=1$ in Eq.~\ref{eqn_collosc}. 

Using this information, the predicted  Fock state sequence fidelity should reach 80\% by ${\bar N}=7$.  This is 15\% higher than we observe in the experiment, and gets slightly worse with increasing $\bar N$.  The source of the additional inefficiency is unknown to us, but we note that our densities,   $5N\times 10^{10}$ cm$^{-3}$, approach peak densities where laser cooling limits are observed due to multiple scattering.  We note that recent results on using Rydberg blockade for single-photon sources \cite{Dudin2012} found a $67\pm10$\% preparation efficiency of the singly-excited many-body state at similar atom densities.


The Q-parameter for deterministic ${\cal N}=1$ schemes studied to date give:  collisional blockade using light assisted collisions, $Q=-0.5$ \cite{DePue1999,Schlosser2002}; this work, $Q=-0.65$, repulsive light-assisted collisions, $Q=-0.91$ \cite{Carpentier2013}; and Mott-insulator samples, $Q=-0.95$ \cite{Bakr2010}.  For ${\cal N}=2$ with $Q=-0.5$,  other methods to date require cooling to quantum degeneracy: the ${\cal N}=2$ shell of  a Mott insulator \cite{Bakr2010}, or controlled spilling from a degenerate Fermi gas \cite{Serwane2011}.  Both achieve $Q<-0.9$.



Our studies of Rydberg-blockade-mediated collective Rabi flopping show that the collective Rabi frequencies very closely follow the predicted $\Omega_N=\sqrt{N}\Omega_1$ dependence.
This, plus our observation of the lack of two atom production in an $A_1B_1$ sequence, strongly imply that the blockade phenomenon is highly effective at rejecting double Rydberg excitations.
We used the collective flopping to produce a strongly sub-Poissonian atom distribution with $Q=-0.65$ in a single trap site.
Extending the protocol to a 2 atom Fock state, we get $Q=-0.5$ and observe 3 cycles of  $N=2$ collective Rabi flopping with no additional dephasing.
Future plans include producing blockaded samples with higher numbers of atoms at lower densities, where possible density dependent mechanisms should be lessened.

\setcounter{figure}{0}
\setcounter{equation}{0}



\acknowledgements{Important contributions at early stages of this work were made by Erich Urban, Thomas Henage,  and Larry Isenhower, and we acknowledge helpful discussions with Klaus M{\o}lmer.  This work was funded by NSF grant \#PHY-1104531 and the AFOSR Quantum Memories MURI.}

\bibliography{/Users/Thad_Walker/Research/thadbibtex/lasercooling}

\end{document}